# Algorithm for Dynamic Fingerprinting Radio Map Creation Using IMU Measurements

**Peter Brída [1], Juraj Machaj [1,\*], Jan Račko [1] and Ondrej Krejcar [2]**

[1] Department of Multimedia and Information Communication Technology, FEIT, University of Zilina,
Univerzitna 1, 01026 Zilina, Slovakia;
juraj.machaj@feit.uniza.sk; jan.racko@feit.uniza.sk; peter.brida@feit.uniza.sk

[2] Center for Basic and Applied Research, Faculty of Informatics and Management, University of Hradec
Kralove, Rokitanskeho 62, 500 03 Hradec Kralove, Czech Republic;
ondrej.krejcar@uhk.cz

\* Correspondence: juraj.machaj@feit.uniza.sk

**Abstract:** While a vast number of location-based services appeared lately, indoor positioning solutions are developed to provide reliable position information in environments where traditionally used satellite-based positioning systems cannot provide access to accurate position estimates. Indoor positioning systems can be based on many technologies, however, radio networks and more precisely Wi-Fi networks seem to attract the attention of a majority of the research teams. The most widely used localization approach used in Wi-Fi-based systems is based on fingerprinting framework. Fingerprinting algorithms, however, require a radio map for position estimation. The paper will describe a solution for dynamic radio map creation, which is aimed to reduce the time required to build a radio map. The proposed solution is using measurements from IMU (Inertial Measurement Unit), which are processed with a particle filter dead reckoning algorithm. Reference points generated by the implemented dead reckoning algorithm are then processed by the proposed reference point merging algorithm, in order to optimize the radio map size and merge similar RPs. The proposed solution was tested in a real-world environment and evaluated by the implementation of deterministic fingerprinting positioning algorithms and achieved results were compared with results achieved with a static radio map. The achieved results presented in the paper show that positioning algorithms achieved similar accuracy even with a dynamic map with a low density of reference points.

## 1. Introduction

With the recent development of location-based services (LBS) in the indoor environment, there is a big demand for the deployment of indoor localization systems [1]. This is mainly because GNSS (Global Navigation Satellite Systems) localization services, which are widely used in the outdoor environment to provide location estimates, cannot provide accurate and reliable localization service in indoor environments. This is mainly related to the properties of GNSS signals that are transmitted from earth orbit and affected by signal attenuation, conditions in the ionosphere as well as multipath propagation. A combination of all these can cause loss of signal or high localization errors in dense urban environments and indoors.

Therefore, alternative positioning solutions have been and are being developed to provide accurate and reliable position estimates in both dense urban and indoor environments. These solutions traditionally utilize data from available enabling technologies to estimate the position of mobile devices or users in the environment. These enabling technologies can be represented by ultrasound [2], cameras [3], light sensors [4], magnetometers [5], MEMS or IMU (Inertial Measurement Unit) [6–8] as well as radio receivers [9]. Each of these technologies has its pros and cons. For example, positioning using cameras can provide high accuracy, however, can have relatively high computation complexity compared to systems based on radio signals [3]. On the other hand, localization based on measurements of magnetic fields can provide high accuracy for moving users with small complexity, however, is almost useless for static positioning as position estimates are based mainly on the classification of changes of the magnetic field when a user moves around the area [10].

When it comes to radio network-based positioning, this can provide position estimates with various levels of accuracy. This is due to the fact that various radio networks use transmit signal on different frequencies with different bandwidths and thus allow the collection of different type of data that can be used to estimate position. For example, UWB (Ultra-Wide Band) technology can provide propagation time ToA (Time of Arrival) measurements thanks to the wide bandwidth of the signal, which is transmitted on high frequencies. Therefore it is possible to use UWB technology to build localization systems based on trilateration and achieve high accuracy even in environments with multipath propagation [11].

An important factor for the deployment of the localization system is the additional cost and complexity required for the operation of the system. The aforementioned UWB localization can provide high performance, however, it required dedicated infrastructure and UWB receivers are not yet commonly implemented in widely used smart devices. Thus, users have to be equipped with tags that are used only for positioning purposes.

The motivation of our work is to build a low-cost positioning system that can provide room-level accuracy without the need for any significant investments into the infrastructure. Therefore, the decision was to set up the localization system on Wi-Fi technology. The main advantage is that Wi-Fi technology can be considered to be ubiquitous and all modern smart devices are able to receive signals from Wi-Fi access points (APs). The localization system is based on fingerprinting framework, since it allows the use of simple Received Signal Strength (RSS) measurements, that does not require any additional modification of devices. Fingerprinting approach has a significant advantage when compared to RSS based trilateration [12] since information about transmit power is not required. Moreover, the effect of shadowing caused by walls and other obstacles as well as multipath fluctuation does not have to be considered when fingerprinting based localization is implemented.

The main drawback of the localization systems based on fingerprinting framework is the need for radio map measurements [13]. The radio map is basically a database of measurements performed on predefined reference points and is used during the position estimation process. The collection of RSS samples at predefined positions is usually a time-consuming process, especially if localization services are to be provided on large scale. One way to avoid time-consuming calibration measurements for radio map is to use complex simulation of radio signal propagation, based either on signal modelling or raytracing algorithms [14]. However, these complex simulations still cannot provide realistic estimates of RSS levels in a dynamic indoor environment and can reduce localization performance [12, 13].

To overcome the problems related to complexity and time required to collect radio map measurements in this paper we proposed a solution for dynamic map creation. The proposed solution is based on a collection of data for a radio map while walking around the localization area. It does not require accurate measurement of the position of each reference point neither static measurements performed at each of the reference points.

The proposed solution is based on the collection of data from IMU and Wi-Fi receiver, which are both implemented in all smart devices available on a market these days. The data from IMU are then processed by a dead reckoning algorithm and particle filter to recover the track of the user. When the track has been successfully recovered the position of reference points with RSS measurements provided by the Wi-Fi receiver can be estimated and stored in the radio map.

The rest of the paper is organised as follows; in Section 2 related solutions used for deployment and experimental evaluation of the systems are described, proposed dynamic map creation solution is described in Section 3, Section 4 describes the experimental setup and discuss achieved results and Section 5 concludes the paper.

## 2. Related Work

### 2.1 Fingerprinting localization

Among localization techniques used for indoor positioning based on Wi-Fi signals, fingerprinting localization has the most attention with a vast number of modifications proposed by many research teams. This is due to the fact that fingerprinting positioning can estimate the position of the user with the use of easily obtained RSS measurements. While traditional distance-based localization methods that use RSS measurements for distance estimation are negatively affected by fluctuations caused by multipath propagation of wireless signals, the fingerprinting-based positioning systems seem to perform much better in environments with dense multipath propagation.

The operation of fingerprinting localization systems can be divided into two separate parts usually referred to as offline and online phases. The offline phase is dedicated to calibration measurements of a radio map, while the online phase describes how the system operates during positioning.

During the offline phase, measurements are traditionally performed at predefined reference points distributed in the area of interest, where localization services will be provided, also referred to as localization area. The measurements on individual reference points are stored in the database widely called a radio map. The principle of the radio map is shown in **Figure 1**. The radio map consists of reference points identification, the position of each reference point, as well as measured data. The position of the reference point is defined by coordinates [$x_{RP}$, $y_{RP}$, $z_{RP}$], where $x_{RP}$ and $y_{RP}$ represent coordinates in 2D cartesian space and $z_{RP}$ defines the floor of the building. The most important information in the radio map is represented by RSS samples measured from all available APs. All the RSS measurements are linked to MAC addresses of APs that transmitted the signal; these are used as identification of individual transmitters during the position estimation process.

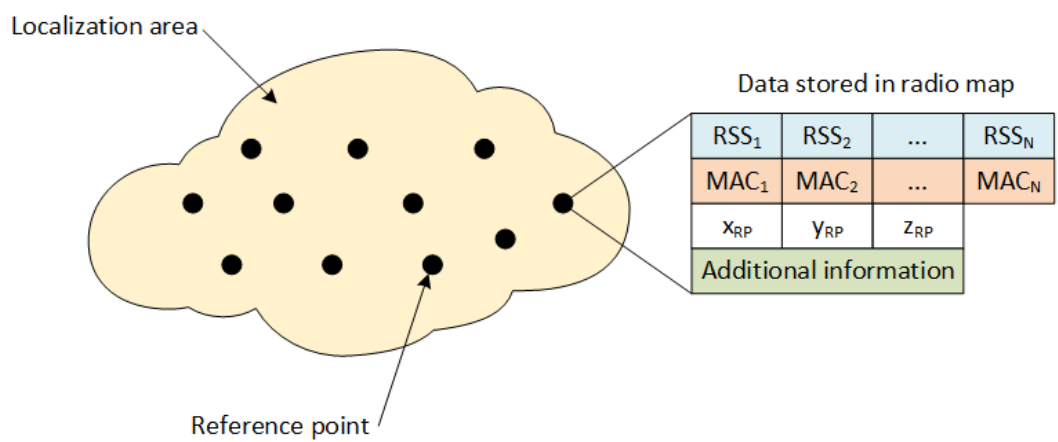

**Figure 1.** Principle of the radio map

Additional information can be included in the radio map as well, this can be an identification of a device that was used for measurements, the orientation of the device during the measurement, etc. This information may help to improve the performance of the localization system.

In the online phase, the main goal is to estimate the position of the mobile device based on actual RSS measurements collected at the unknown position. This can be done by algorithms that can be divided into three main categories – deterministic approach, probabilistic approach and based on machine learning techniques.

Recently, machine learning algorithms based on neural networks and support vector machines are becoming extremely popular in different areas of signal processing and data analysis. Therefore their use in localization systems is gaining a lot of attention as well [14, 15].

In the probabilistic approach, the position of the mobile device is assumed to be a random vector [19]. In this case, the localization candidate is commonly chosen based on the highest posterior probability of the location candidates. If RSS measurements are represented by a vector $S$ and $\gamma_i$ represents position for $i$-th reference point, then the decision rule based on Bayes' theorem is given as follows:

$$P(\gamma_i|S) = \frac{P(S|\gamma_i)P(\gamma_i)}{P(S)},$$ (1)

where posterior probability $P(\gamma_i|S)$ is described as a function of likelihood $P(S|\gamma_i)$, prior probability $P(\gamma_i)$ with uniform distribution if no prior data are available, and observed evidence $P(S)$ given by:

$$P(S) = \sum_i P(S|\gamma_i)P(\gamma_i).$$ (2)

Other likelihood functions can also be used as can be seen in [20].

On contrary, in the deterministic framework, the state of the mobile device is assumed to be a non-random vector. The main idea is that position of the mobile device can be estimated since RSS values depend on the position of the mobile device. Therefore, the position of the mobile device $\hat{x}$ can be estimated based on a direct comparison of RSS values measured by the mobile device with values stored in the radio map database. In this case, the estimator can be written as:

$$\hat{x} = \frac{\sum_{i=1}^{M} \omega_i \cdot \gamma_i}{\sum_{i=1}^{M} \omega_i},$$ (3)

where $M$ is the number of reference points in the radio map database $\omega_i$ represents nonnegative weight assigned to the $i$-th reference point with position $\gamma_i$. Weights $\omega_i$ can be calculated using different metrics [21], however, the most commonly used metric is the Euclidean distance $d_E$ given by:

$$d_E = \sqrt{\sum_{k=1}^{N}(a_k - b_k)^2},$$ (4)

where $N$ is the number of elements in the RSS vector, i.e., the number of APs detected by the Wi-Fi receiver, $a_k$ represents $k$-th element of vector $A$ and $b_k$ represents $k$-th element of vector $B$.

The positioning algorithm that uses estimator (3) to estimate position using $K$ highest weights is called Weighted K-Nearest Neighbours (WKNN) [22], if $K$ highest weights in the estimator (3) are set to 1 and other weights set to 0 represents KNN (K-Nearest Neighbours) algorithm. If the position of the mobile device is estimated using only a single highest weight the algorithm becomes NN (Nearest Neighbour) algorithm [23].

In general, KNN and WKNN provide more accurate position estimates, especially when K is set to 3 or 4 since it is possible to estimate the position of the mobile device more precisely. However, the NN algorithm can achieve an accuracy similar to KNN and WKNN algorithms when the density of the radio map is high enough [20]. This is due to the fact that the NN algorithm can only estimate positions on reference points, thus the minimum error is given by the distance between them.

*2.2 Improvements of the radio map*

Since the radio map is required for fingerprinting-based localization, a lot of attention was aimed at modifications of the radio map collection process or dynamic updates of the radio map. Most of the proposed solutions can be divided into two main categories. The first category consists of solutions that help to increase radio map density by the implementation of interpolation algorithms to estimate fingerprints on additional reference points from the manually measured radio map. The second category of solutions is aimed at online radio map updates using measurements taken during the localization phase, either from a localized device or set of sensors placed in the localization area.

The interpolation methods are used to increase radio map density, these solutions are based on assumption that the smaller distance between reference points the better resolution of the localization system and thus it should be possible to achieve higher localization accuracy. The first solutions for interpolation of reference points in sparse radio maps were based on linear interpolation [24], Radial Basis Function (RBF) [25], Kiring interpolation [26] as well as Inverse Distance Weighting (IDW) [27]. A more advanced solution referred to as VORO was proposed by Lee and Han [28] and is based on Log-Distance Path Loss. The interpolation was performed in two phases. In the first phase, the positions of APs were estimated. In the second phase parameters of signal, fading was estimated for each Voronoi cell, while fading of walls and obstacles was taken into the account. The parameters of signal fading were then used to estimate RSS on interpolated reference points. Based on experimental results the VORO solution achieved significantly better results than IWF and RBF based interpolations.

Another solution for interpolation of radio map was proposed by Khalajmehrabadi et al. in [29], in this case, interpolation was reformulated as a minimization problem known as the Least Absolute Shrinkage and Selection Operator (LASSO). The main advantage is that the proposed algorithm interpolates signals from randomly selected reference points and new samples are estimated for each AP individually, thus seems to be more robust.

Application of the Kiring interpolation technique on radio map re-emerged recently in applications related to 5G communication [30] and V2X (Vehicle to everything) [31] with distributed implementation. In these cases, it is required to estimate the power of the received signal for multiple nodes in the area. Kiring interpolation can achieve reasonably good performance since the implementation is mainly aimed at vehicle communication in the outdoor environment without any significant shadowing caused by obstacles between the transmitter and individual receivers.

On the other hand, algorithms in the second category aimed at online radio map updates are based on assumption that the radio map is changing over time, due to changes in the environment and thus by regular updates of radio map it will be possible to improve localization accuracy. The changes in the environment might be caused by variation of temperature, humidity, movement of obstacles or changes in the infrastructure, i.e., furniture and equipment, reconstruction of the building, replacement or upgrade of some transmitters in the area [32]. These solutions are based on machine learning [33] and regression methods [34].

Xu et.al proposed an online radio map update scheme based on the marginalized particle gaussian process in [35]. The proposed solution utilizes crowdsourced fingerprints to update the radio map, while the position of online measurements is estimated using the existing radio map. The advantage of the solution lies in recursive processing of the measurements until the location is aligned with the radio map.

Huang et. al. [36] implemented a marginalized particle gaussian process in combination with pedestrian dead reckoning and Wi-Fi based localization for alignment of measurements with the existing radio map. Results show that implemented solution can provide better localization performance than the Gaussian process regression approach, however, requires significantly more processing power. Thus, the online update is not automatic but has to be scheduled so it will not disrupt localization service.

The online radio map update using measurements using data from the fixed nodes was proposed by Batalla et. al. [37]. The authors proposed to use a number of nodes to monitor changes in radio signal propagation. These devices are placed on some of the reference points. Measurements performed by the fixed nodes are used to update the offline radio map. The disadvantage of the system is a requirement for the implementation of fixed nodes and the fact that the number of nodes will increase significantly when the localization area consists of a large number of rooms since conditions in different rooms will change independently, unlike in large open spaces like industrial halls where some correlation between neighbouring reference points can be expected without constraints.

All the solutions described above can help to improve the performance of the localization system based on fingerprinting framework. However, all these solutions require initial calibration measurements for the radio map, which are time-consuming. Some of the solutions require additional investment into infrastructure, to collect data for online radio map updates.

The dynamic online calibrated radio map proposed in [38] helps to construct the radio map of the localization area and provide online updates of the radio map database based on the measurements performed by APs implemented in the area. This solution is quite promising; however, it is based on the assumption that most of the APs in the area are capable to perform RSS measurements and send them to the localization server. This might be true in the case when the network in the whole localization area (building) has a single administrator, unfortunately, this is not always the case.

In this paper, we focus on the low-cost solution, focused on the reduction of effort required for initial calibration measurements, which is required for most of the solutions presented above and does not require access to network infrastructure.

### 2.3 Dead reckoning with particle filter

The dead reckoning positioning using data form low-cost IMU has attracted a lot of attention lately[8]. In the previous work [39] we described the implementation of the dead reckoning algorithm with a particle filter. The dead reckoning algorithm is utilizing data from IMU to reconstruct the track of the user. The implementation can be divided into three parts, in the first part algorithm for step detection is implemented, the second part consists of an algorithm for heading angle estimation and the third part is represented by a particle filter algorithm that uses map information to improve position estimates. The block diagram of the particle filter pedestrian dead reckoning (PF-PDR) is shown in **Figure 2**.

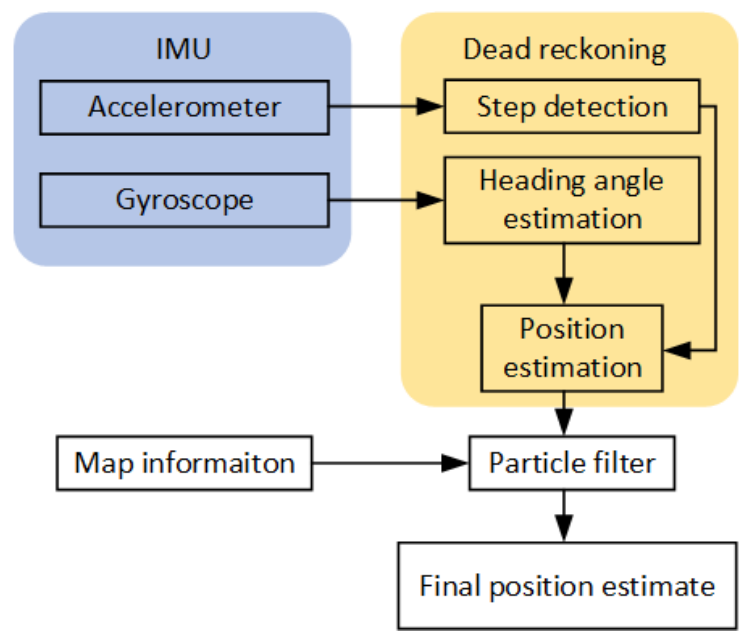

**Figure 2.** Block diagram of particle filter pedestrian dead reckoning

In the implemented dead reckoning solution the first task is to detect the step of the user [40]. For this purpose, data from the accelerometer implemented in IMU are used. The significant pattern of the step can be found in the vertical axis; therefore, a vertical component of acceleration has to be estimated accurately. This cannot always be possible, especially when low-cost accelerometers are used for data collection, thus the norm of acceleration can be used to detect the step. The norm of acceleration is given by:

$$a(t) = \sqrt{a_x^2(t) + a_y^2(t) + a_z^2(t)} - g,$$ (5)

where $g$ is gravitational acceleration and $a_x$, $a_y$, $a_z$ stand for measured acceleration in all three axes of the accelerometer implemented in IMU [41]. The norm can be used for step detection using either peak detection, frequency analysis of the signal or zero-crossing method, which was implemented in our case. In theory, each step has a different length, and this should be taken into the account during the implementation of the dead reckoning algorithm, however, in this case, a fixed step length of 0.75 m was used. This setting was tested in [39] and provided reasonably accurate position estimates thanks to the implementation of a particle filter.

The second part of the dead reckoning algorithm is based on heading estimation, which can be done by integration of angular velocity measured by the gyroscope and given by:

$$\omega_b(t) = \left(\omega_x(t), \omega_y(t), \omega_z(t)\right),$$ (6)

where $\omega_x, \omega_y, \omega_z$ are angular rotations measured for each axis in the body frame of the IMU. Then attitude of the IMU can be then represented by direction cosine matrix $\boldsymbol{C}$, which is the rotation matrix given by:

$$\boldsymbol{C} = \begin{bmatrix} \cos\theta\cos\Psi & \cos\theta\sin\Psi & -\sin\theta \\ \sin\varphi\sin\theta\cos\Psi - \cos\varphi\sin\Psi & \sin\varphi\sin\theta\sin\Psi + \cos\varphi\cos\Psi & \sin\varphi\cos\theta \\ \cos\varphi\sin\theta\cos\Psi + \sin\varphi\sin\Psi & \cos\varphi\sin\theta\sin\Psi - \sin\varphi\sin\Psi & \cos\varphi\cos\theta \end{bmatrix},$$ (7)

where $\varphi, \theta, \Psi$ represent Euler angles roll, pitch and yaw, respectively. In order to track the orientation of the IMU the rotation matrix has to be updated all the time. The updated matrix $\boldsymbol{C}(t + \Delta t)$ can be calculated as follows:

$$\boldsymbol{C}(t + \Delta t) = \boldsymbol{C}(t)\left(\boldsymbol{I} + \frac{\sin\sigma}{\sigma}\boldsymbol{B} + \frac{1 - \cos\sigma}{\sigma^2}\boldsymbol{B}^2\right),$$ (8)

where $\Delta t$ is the sampling interval, $\boldsymbol{I}$ is 3-by-3 identity matrix, $\sigma = |\Delta t\omega_b|$ and

$$\boldsymbol{B} = \begin{bmatrix} 0 & -\omega_z\Delta t & \omega_y\Delta t \\ \omega_z\Delta t & 0 & -\omega_x\Delta t \\ -\omega_y\Delta t & \omega_x\Delta t & 0 \end{bmatrix}.$$ (9)

Afterwards, it is possible to calculate yaw angle $\Psi$ from the updated rotation matrix, which actually represents the heading of the user:

$$\Psi = \arctan(C_{2,1}, C_{1,1}).$$ (10)

In the last step of the dead reckoning algorithm the position of the user can be estimated as follows:

$$\begin{bmatrix} P_{x_k} \\ P_{y_k} \end{bmatrix} = \begin{bmatrix} P_{x_{k-1}} + l_k\sin\Psi_k \\ P_{y_{k-1}} + l_k\cos\Psi_k \end{bmatrix},$$ (11)

where $P_{x_k}$ and $P_{y_k}$ represent a position on x-axis and y-axis in step $k$, $l_k$ stands for step length and $\Psi_k$ represents the heading angle in step $k$.

To achieve optimal combination information from various sources Bayesian filters are widely used, unfortunately, Bayesian filters work well only with linear models. Since the localization process is nonlinear, an approximation of the Bayesian filter must be implemented. A particle filter is a popular approximation, where posterior state distribution is approximated using particles. Moreover, the advantage of the particle filter is that representation of particles as standalone points can easily be combined with information about the area in the map-matching process [42].

The particle filter is operating in prediction and update steps. During the prediction step, the number of particles $x^{(p)}$, $p = 1, \ldots, N$ is generated from proposal distribution in a time step $t$:

$$x_t^{(p)} \approx \pi \left( x_t^{(p)} | x_{t-1}^{(p)}, \; y_{1\ldots t-1} \right), \tag{12}$$

where $y_{1\ldots t-1}$ stands for measurements one step before $t$. It is assumed that states establish a Markov model, this current state $x_t$ depends solely on the previous state $x_{t-1}$.

During the uprate phase, it is required to recalculate weights according to the likelihood of observation, so the weights are given as follows:

$$w_i^p = w_{i-1}^p \frac{p\left(y_t | x_t^{(p)}\right) p\left(x_t^{(p)} | x_{t-1}^{(p)}\right)}{q\left(x_t^{(p)} | x_{t-1}^{(p)}, y_i\right)}. \tag{13}$$

After the update, the weights are normalized. During the operation of the particle filters, just a few particles will be assigned all calculated weights. Over time propagation of particles with low weights has a negative impact on posterior distribution, which is referred to as degradation. In order to avoid degradation, resampling is required [43]. The implemented solution performed resampling when the number of active particles was less than $N/5$.

## 3. Proposed Solution

In order to remove the biggest drawback of fingerprinting positioning systems, energy and time required for the construction of a radio map, we have proposed a solution for dynamic radio map creation. The proposed solution utilizes IMU data and dead reckoning using the particle filter presented in [39]. In traditional fingerprinting solutions, the radio map is created either by performing measurements on predefined spots or by complex simulations of radio propagation [44]. However, simulations cannot provide realistic results even when a large number of complex factors affecting radio signal levels are considered [45]. Therefore, systems with radio maps based on real-world measurements traditionally outperform systems based on artificial radio maps [16].

Our proposed dynamic map creation solution is aimed at the reduction of radio map measurements complexity since the solution can provide an automatic collection of RSS fingerprints for radio map. The proposed solution is based on RSS measurements during a walk through the localization area. Data are collected using the developed Sensor Reader app which allows collecting data from both IMU and Wi-Fi receiver at the defined sampling rates. In the application, the sampling rate for IMU data was set to 100 Hz and for Wi-Fi signals, the measurements of RSS from all surrounding APs were performed every 5 s in order to provide reasonable separation for reference points in the radio map.

In the second step, the data collected by the user during the walk around the area is processed by the dynamic map creation algorithm. The dynamic map creation algorithm use data from IMU, which are processed by the PF-PDR algorithm in order to reconstruct the trajectory of the user. Furthermore, reference points are defined on the trajectory on time steps corresponding to Wi-Fi measurements.

The last step of the proposed solution is the process of reference point merging. This process is used to combine measurements from the reference points with similar position or neighbouring reference points with extremely similar RSS values. The flowchart of the dynamic map creation, as well as the Reference Point (RP), merge process is shown in **Figure 3.**

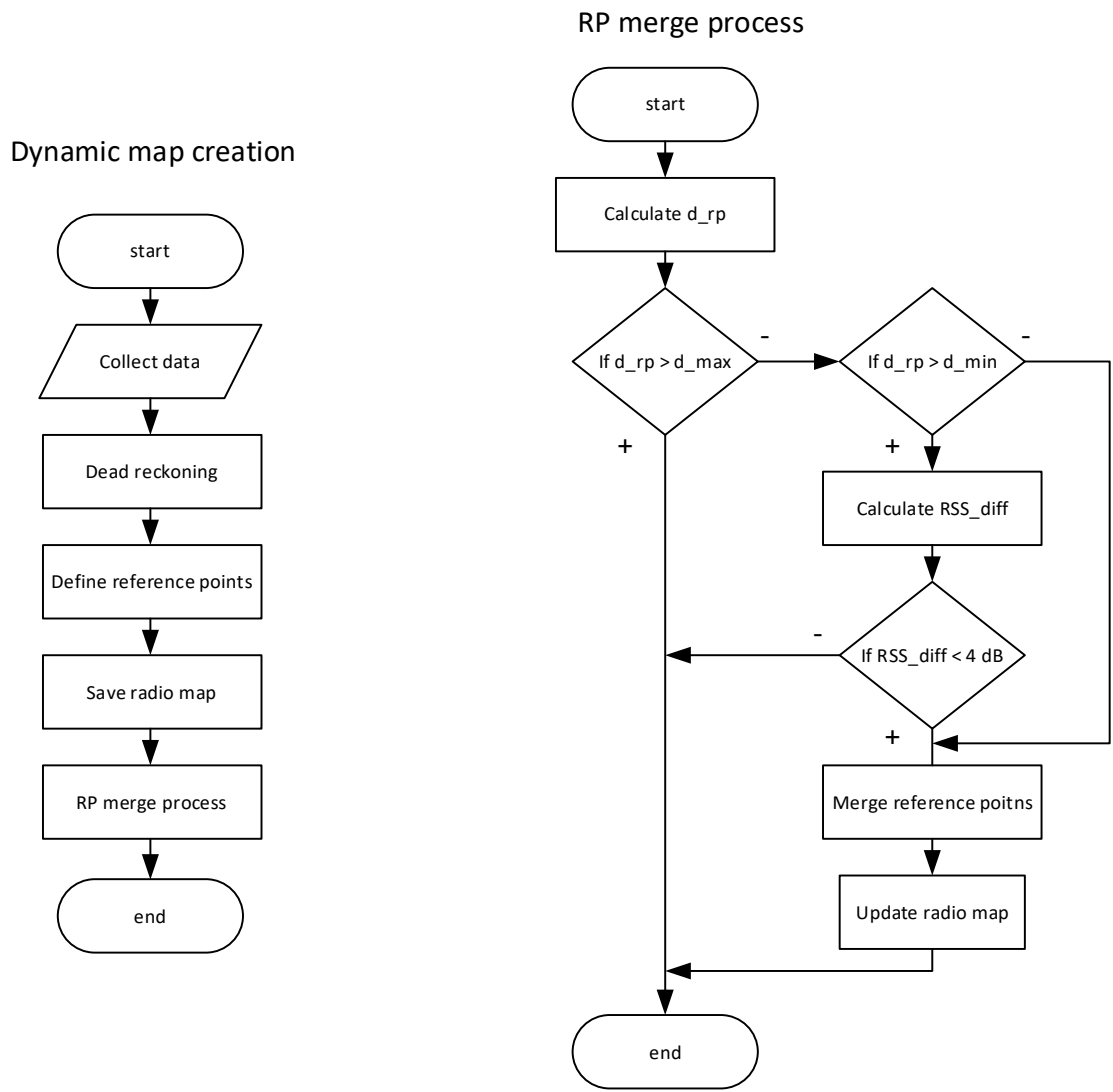

**Figure 3.** Flowchart of the proposed dynamic map creation solution and RP merge process.

Since the time interval for RSS measurements had a fixed value of 5 s it is possible to estimate positions of individual reference points on which the radio map measurements were performed. It could happen that RSS measurements were performed during the same period as the user made a step which was detected by a dead reckoning algorithm, in such case the measurements are linked to the posterior position of the user.

However, since the user can move freely around the area while performing the radio map measurements it can happen that some RSS measurements are performed at reference points that are close to each other. In order to achieve an even distribution of reference points in the radio map, the RP merge process was proposed. The process will run for each combination of neighbouring reference points and in the first step calculate the physical distance between them. If the distance is smaller than *d_max*=4 m the merging process will continue. The distance is then compared to *d_min* and if it is smaller the neighbouring RPs will be merged without a comparison of the RSS values. However, if the distance is between *d_max* and *d_min* the algorithm will continue with a comparison of RSS values on the neighbouring reference points in order to decide about merging.

The RSS comparison is based on the computation of the mean RSS difference between the neighbouring reference points denoted as $RSS_{dif}$, the value is given as follows:

$$RSS_{dif} = \frac{1}{n} \sum_1^n |RSS_n^1 - RSS_n^2|, \tag{14}$$

where $n$ is the number of unique APs detected on both neighbouring points, defined by the number of unique MAC addresses, $RSS_n^1$ and $RSS_n^2$ represent RSS samples measured from AP with $n$-th MAC address at the first and the second neighbouring node, respectively. The threshold for the merging of neighbouring nodes was set to 4 dB. When the $RSS_{dif}$ is higher than the threshold the neighbouring nodes are considered to represent significantly different conditions in the radio channel [46], therefore the merging of reference points is not performed and both reference points will be considered in the final radio map database.

In case that all required conditions were met in the algorithm, i.e. the physical distance between RPs is below $d\_min$ or the average RSS difference $RSS_{dif}$ is above 4 dB, the merging of reference points will be performed.

In the merging process, the position of the merged reference point is defined as the average of positions of the original reference points. The RSS values for the merged reference point are calculated as average values of RSS samples collected for both original reference points. In a case when some APs are detected only on one of the original reference points values are considered to be -100 dBm, which represents the sensitivity of a Wi-Fi receiver.

## 4. Achieved Results and Discussion

### 4.1 Proof of concept of dynamic radio map creation

In order to test and prove the feasibility of our solution, we have performed initial testing of the system for the dynamic radio map construction at the Department of Multimedia and Information-communication technology of the University of Zilina. The walked path recovered from the IMU data processed by particle filter based dead reckoning algorithm is shown in **Figure 4**, red circles in the figure represent positions of reference points created by the dynamic radio map creation solution.

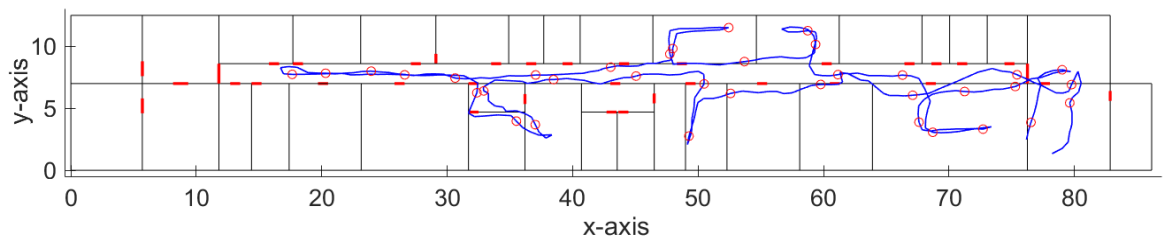

**Figure 4.** Track recovered by the dead reckoning algorithm with detected reference points.

We have also evaluated the accuracy of the implemented dead reckoning algorithm, which is an important parameter since it has an impact on the accuracy of reference points position. The localization error of the dead reckoning algorithm with particle filter is presented in Table 1. The localization error was estimated as a distance between the recovered track and the reference track. The ground truth position was estimated thanks to time stamps assigned at known positions, during the experiment constant movement speed was considered. Therefore, ground truth positions can be defined on a line between two known points with desired time steps. Thus, it was possible to link position estimates in individual time steps with ground truth position estimates and calculate localization error for each time step.

**Table 1.** Localization error of the dead reckoning algorithm.

| Localization error [m] | | | | |
|---|---|---|---|---|
| minimum | median | mean | 90% | maximum |
| 0.007 | 0.43 | 0.59 | 1.64 | 2.44 |

From the results presented in the table, it can be seen that the average localization error achieved by the implemented algorithm was around 0.6 m. Moreover, it is well known that localization error achieved by the dead reckoning algorithm is increasing with time due to the integration of error from sensors, therefore this can be reduced if measurements are performed on multiple shorter tracks starting from a known position. We assume that the average localization error of 0.6 m will not have a significant impact on localization accuracy since it is significantly lower than the resolution of the radio map, which is in our case assumed to be 2 m. Therefore, in case the distance between the neighbouring dynamic reference points will be less than 4 m the reference points will be processed by the RP merging algorithm described in the previous section. The algorithm will automatically merge points with a distance less than 2 m, and perform a similarity check for neighbouring RPs with a distance up to 4 m.

With the further analysis of achieved results, we have found out that in some cases the track estimated by the implemented dead reckoning algorithm, although with reasonably small error, is still crossing between the rooms thru walls as can be seen in **Figure 5**.

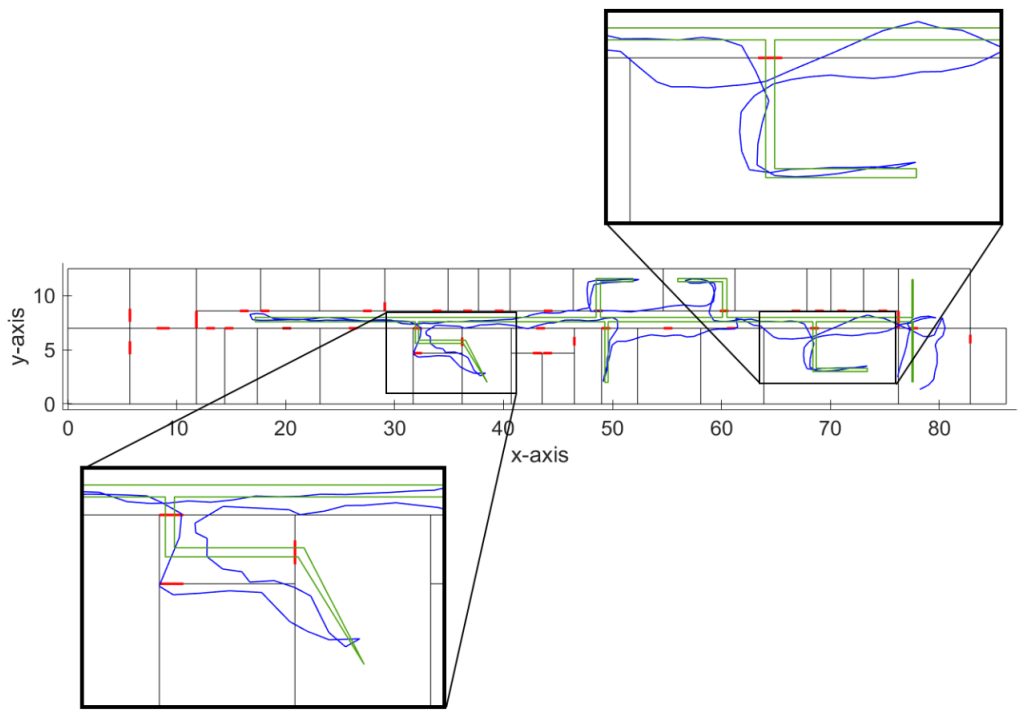

**Figure 5**. Detailed view of errors of the recovered track.

These localization errors might be a problem in cases when IMU data are used for real-time navigation, however, in an application focused on dynamic radio map construction this can be neglected since the localization error introduced by this phenomenon will still be reasonably low. Moreover, these errors mainly occurred when moving thru the corridors, where multiple neighbouring reference points are expected to be merged. This is because the user usually travels thru corridors multiple times, therefore more reference points are expected to appear in this area. Moreover, if corridors are passed multiple times by the user with different localization error, it can be expected that part of the localization error can be mitigated by the RP merging algorithm. Since these errors are likely to occur in different directions. Therefore, the average position of merged reference points can be corrected. We assume this problem of the dead reckoning algorithm can be further reduced by the modification of weights in the used particle filter, which will be investigated in the future.

Since in fingerprinting localization system the position is estimated based on RSS comparison it is important to have accurate RSS measurements on each reference point. Therefore, in order to prove the feasibility of the dynamic map creation, we have performed some static measurements on the positions of reference points and compared the RSS samples. The RSS values in static measurements were averaged from the 20 samples since this process helps to reduce RSS fluctuations and is used during static radio map measurements. The difference of RSS calculated from static and dynamic measurements at 7 reference points for all detected APs is presented in **Table 2**. The values in the table are based on a comparison of RSS samples from 86 APs, with signals from the same AP detected at different RP treated as unique samples. The number of APs detected for RPs is not the same across the map, therefore selected RPs had between 5 and 16 detected signals, representing RPs with poor as well as good coverage.

**Table 2.** Difference between static and dynamic RSS values.

| Difference of RSS values [dB] | | | | |
|---|---|---|---|---|
| minimum | median | mean | 90% | maximum |
| 0.024 | 1.99 | 2.057 | 5.123 | 12.126 |

From the table, it can be seen that on average the difference between RSS values from dynamic and static measurements is just above 2 dB, with 90% of differences than 5.2 dB, which might be caused by signal fluctuations. However, the maximum difference is above 12 dB, which seems to be high enough to cause localization errors. However, a number of such high RSS differences was less than 10% of all samples. It is important to note here, that RSS measurements can also be affected by the orientation of the mobile device as well as attenuation caused by the human body. In some cases, the difference of RSS caused by the human body can be up to 10 dB for signals in the ISM band [47], which is the frequency band used for 802.11b/g/n Wi-Fi signal transmission.

Based on the results we can conclude that the proposed solution for dynamic radio map creation was able to provide RSS measurements with reasonable accuracy, while significantly reduce the time required to perform the measurements required for radio map. When compared to static measurements of radio map it was possible to reduce the required time by 90%, therefore making it possible to perform multiple dynamic measurements and still reduce the required effort significantly.

### 4.2 Localization performance

We have also evaluated the performance of the localization system with the use of dynamic radio map creation. The radio map used in the experiment was created using the dynamic map creation algorithm, in total 66 reference points was created. The total number of unique detected APs was 97. The number of APs detected on a single reference point in the radio map was between 5 and 32, with an average of 16 APs per reference point.

To test the feasibility of a dynamic radio map we have performed 38 position estimates at points that did not correspond to locations of reference points in the radio map. Localization was performed using NN, KNN and WKNN algorithms and achieved localization error is presented in **Table 3**. In the table results achieved for the static map are presented as well. These results were achieved with the static radio map, with calibration measurements taken at reference points placed in a grid with 2 m separations. In the case of the static map, the position of the mobile device was estimated at 25 positions.

**Table 3.** Localization error.

| Localization error [m] | | | | |
|---|---|---|---|---|
| Algorithm | Minimum | Mean | Median | Standard deviation |

| | | | | | |
|---|---|---|---|---|---|
| **Dynamic map** | **NN** | 1.01 | 3.51 | 3.01 | 1.97 |
| | **KNN** | 0.43 | 4.05 | 3.39 | 2.64 |
| | **WKNN** | 0.59 | 3.77 | 2.9 | 2.46 |
| **Static map** | **NN** | 0 | 2.91 | 3 | 2.28 |
| | **KNN** | 0.33 | 2.75 | 2.43 | 1.33 |
| | **WKNN** | 0.24 | 2.72 | 2.42 | 1.39 |

The results presented in the table for KNN and WKNN algorithms were achieved with *K* = 3, as this setting should in theory provide the optimal localization performance. However, in case, when the dynamic radio map was used, it seems to perform worse than the NN algorithm. This is caused by the small density of reference points in the radio map. For further analysis, the CDF of localization errors achieved by localization algorithms with a dynamic radio map is presented in **Figure 6**.

It is important to note here that all algorithms performed better with the static radio map, which was expected since the radio map had a higher density of reference points. Interestingly enough, there was no significant difference in the median localization error for the NN algorithm, since in both cases the median error was 3 m. While the difference in median value was just around 0.5 m for the WKNN algorithm and approximately 0.9 m for the KNN algorithm. On the other hand, the difference in the mean localization error ranges from 0.6 m for the NN algorithm up to 1.3 m for the KNN algorithm.

The lack of difference in median error, which is given by 50 percentiles of all achieved errors, for the NN algorithm can be caused by the fact that in some localization points the distance to the nearest reference point in the radio map was similar in both scenarios. However, due to the lower density of the dynamic radio map, the average error of the NN algorithm was negatively affected by higher errors at higher percentiles.

Based on these results we can conclude that the dynamic radio map can provide reasonably accurate position estimates, while significantly reducing the effort required to set up a fingerprinting localization system.

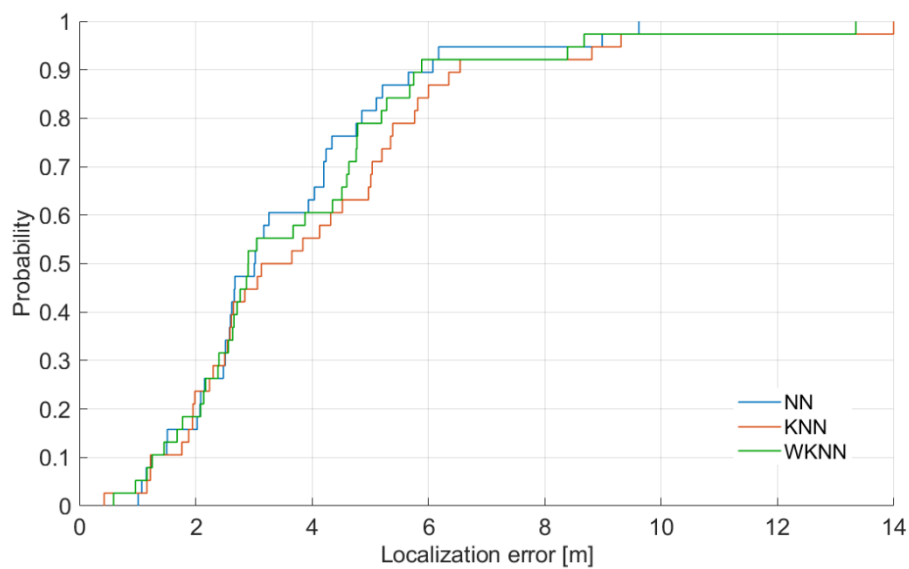

**Figure 6.** CDF of achieved localization errors.

From the data presented in the figure, it is clear that under the given conditions the NN algorithm outperformed both KNN and WKNN algorithms. However, it is clear that the difference between the localization results provided by all three algorithms is quite small. Nevertheless, the worst performance was achieved by the KNN algorithm. This is caused by the fact that in KNN algorithms selected *K* reference points to contribute to the final position estimate with the same weight which is not ideal in the case when the mobile device is not placed in the centre of gravity of the selected reference points.

It can be seen that the minimum error is between 0.4 m and 1 m, which is expected due to the limited resolution of the radio map in the experiment. It can be concluded that in 40% of location estimates the performance of all algorithms was almost the same. The difference in positioning performance is clearer for 50% error where the KNN algorithm starts to perform significantly worse than other algorithms. It can be concluded that WKNN and NN algorithms achieved localization error below 6 m in 90% of position estimates, while the KNN algorithm achieved an error smaller than 6 m in 85% of estimates.

Since the radio map of the localization area is sparse and thus might be causing higher localization errors in the case when a larger number of reference points is used in the position estimation process, we also investigated the impact of the different number of considered reference points, i.e. different *K* value, on localization accuracy, achieved medial localization errors can be seen in **Figure 7**.

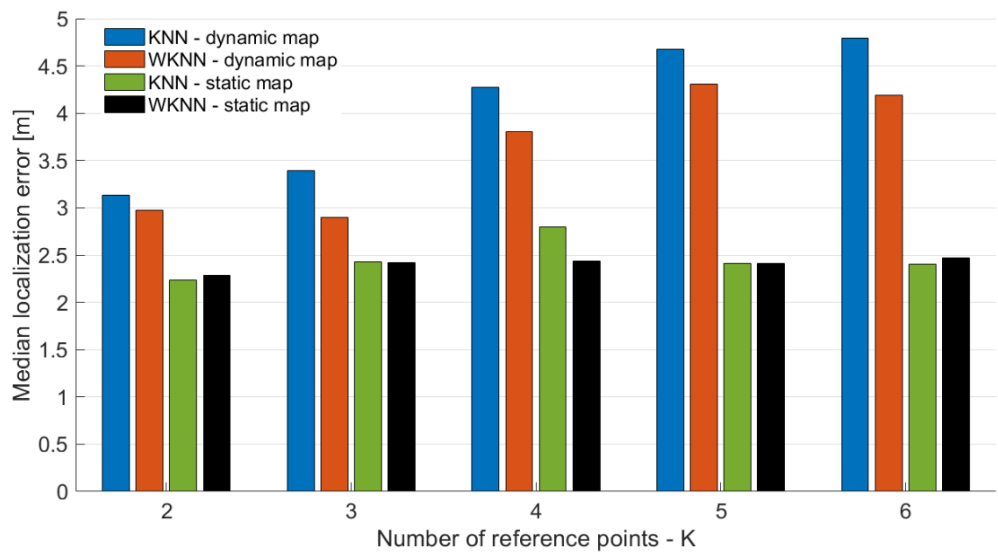

**Figure 7.** Impact of the number of reference points used for position estimation.

From the results presented in the figure, it can be seen that *K*, representing the number of reference points used in KNN and WKNN algorithms, has an impact on localization accuracy. This holds for both static and dynamic map, however, due to the low density of reference points in the dynamic radio map, the effect is different than in localization with a static map. When the dynamic radio map was used the higher value of *K* leads to a higher localization error since reference points further from the actual position are selected and considered in the localization process. On contrary, it can be seen that *K* has a smaller impact on localization error with static radio map.

When the median localization error of KNN and WKNN algorithms with dynamic radio map is compared to the error achieved by the NN algorithm with dynamic radio map it can be concluded that NN outperformed KNN algorithm for any value of *K* since the KNN algorithm achieved the lowest error with *K* = 2 and the error was 3.13 m, while the NN algorithm achieved the median error of 3.02 m. On the other hand, the WKNN algorithm achieved a lower median error than the NN algorithm for both *K* = 2 and *K* = 3.

From the results, it can be seen that the best localization error was achieved for *K* = 2 for both dynamic and static radio maps. This is in contrast with the assumptions and results presented in previously published papers [17, 20]. The fact that *K* = 2 achieved the lowest localization error for almost all cases, except WKNN with dynamic radio map, might be given by the fact that experiments were performed in the building with relatively small offices and narrow corridors, therefore each room had only 2 or 4 reference points.

This caused that with higher *K* some reference points were selected in incorrect rooms and therefore had a negative impact on localization error.

Interestingly enough the best results for each KNN and WKNN were achieved with the different number of considered reference points. The lowest median error was achieved for *K* = 2 and *K* = 3 for both KNN and WKNN algorithms, respectively. The low optimal value of *K* is given by the low density of the radio map, however, it is interesting to see that even with sparse reference points in the radio map the achieved localization accuracy is close to the accuracy achieved with the radio map with reference points in the grid with 2 m spacing that requires significantly higher effort and consumes significantly more time to perform calibration measurements for radio map.

## 5. Conclusions

In the paper, the solution for dynamic radio map collection was introduced. The proposed solution is based on simultaneous measurements of RSS from Wi-Fi networks and the collection of IMU data. The IMU data is processed by the dead reckoning algorithm with particle filtering, which helps to reduce the localization error of the recovered track. The proposed solution was tested in a real-world environment. The mean localization error of the recovered track was less than 0.6 m with a maximum error of approximately 2.5 m.

The algorithm for reference points merging that helped to process radio map data was introduced since some reference points generated form the PF-PDR algorithm were too close to each other and had too similar RSS measurements. Moreover, generated reference points were compared with static measurements. From the comparison, it can be concluded that the dynamic points had reasonably accurate RSS measurements with a mean RSS difference from static measurements around 2 dB. Therefore, we concluded that the dynamic radio map can be suitable for positioning.

In order to provide proof of concept, the feasibility of the radio map was tested in the localization system with NN family algorithms. Interestingly enough the achieved mean localization error was similar to results achieved in our previous experiments with a much denser radio map created using static measurements. Moreover, the NN algorithm achieved the lowest mean localization error and the lowest standards deviation among all three algorithms. This might be caused by the extremely low density of reference points in the radio map. Therefore, the distance between the reference points selected by KNN or WKNN might have been too high resulting in higher localization errors. This was also proved by the experiment aimed at evaluation of the impact of the number of reference points used for position estimation in KNN and WKNN algorithms, where it can be seen that with higher *K* the localization error of the KNN algorithm is increasing significantly, which is not the case when radio map with a higher density of reference points was used.

**Supplementary Materials:** No supplementary materials.

**Author Contributions:** Conceptualization, J.M., J.R. and P.B.; methodology, J.M., J.R., P.B. and O.K.; investigation, J.R., J.M. and P.B.; writing and editing, data curation, visualization, formal analysis, J.M.; supervision, P.B.; funding acquisition, P.B., J.M. and O.K. All authors have read and agreed to the published version of the manuscript.

**Funding:** This work has been partially supported by the Slovak VEGA grant agency, Project No. 1/0626/19 "Research of mobile objects localization in IoT environment" and the European Union's Horizon 2020 research and innovation programme under the Marie Skłodowska-Curie grant agreement No 734331.

**Institutional Review Board Statement:** Not applicable.